# Phase-field model of ion transport and intercalation in lithium-ion battery

Pavel E. L'vov[1,2*], Mikhail Yu. Tikhonchev[1], and Renat T. Sibatov[3]

[1]*Ulyanovsk State University, Ulyanovsk, Russia, 432017*
[2] *Institute of Nanotechnology of Microelectronics of the Russian Academy of Science, Moscow, Russia, 119991*
[3] *Moscow Institute of Physics and Technology, Dolgoprudny, Moscow, Russia, 141701*
[*]Corresponding author (E-mail: LvovPE@sv.uven.ru)

## Abstract

The unified 3D phase-field model for description of the lithium-ion cell as a whole is developed. The model takes into account the realistic distribution of particles in porous electrodes, percolative transport of ions, and difference in size of solute and solvent molecules. We use spatially dependent interaction and dynamic parameters that are considered as a function of the order parameter determining the space and size distribution of particles in nanostructured porous electrodes. The model describes the dynamics of ions in a battery at the constant value of overpotential across the electrode/electrolyte interface. The electrochemical reaction is naturally determined by the chemical potential difference at the interface. The proposed model is applied to simulate charging and discharging process in 3D lithium-ion cell with porous cathode and anode characterized by the realistic size distribution of electrode particles. We demonstrate that the condition of constant overpotential provides nearly constant current density in the battery. The obtained results indicate non-diffusive motion of the concentration front during charging or discharging, nonuniform intercalation flux over the surface of electrode particles, and violation of the equipartition of electric current density over the electrode surface. All these facts are not taken into account by most commonly used models.

## 1. Introduction

The explosive growth in energy storage devices has made lithium-ion batteries (LIBs) attractive for the extensive experimental and theoretical studies. LIBs are characterized by high power density, long life, low self-discharge, and exhibit no memory effect.[1,2] These advantages provide a wide employment of LIBs in portable electronics. The intercalation and transport of ions are the key processes that underlie the functioning of electrochemical systems including LIBs.[2,3] Therefore, their understanding is essential for improvement of the battery performance characteristics.

The battery functioning can be analyzed using the electrochemical system theory[2-6], which underlies the widely-used single particle (SP) and pseudo-two-dimensional (P2D) models.[4-12] In accordance with these models, the process of insertion or extraction is studied under galvanostatic regime by the solution of the diffusion equation in spherical electrode particle with the second-type boundary condition. The interfacial current density is considered to be constant over the whole particle interface. This boundary condition is introduced in terms of the well-known Butler-Volmer (BV) relation.[2-12] The transport of ions in electrolyte is usually described by the one-dimensional diffusion equation including the source term following the galvanostatic regime of electrode particle charging/discharging.[2-12] The porosity of electrodes is described by the empirical Bruggeman coefficient determining the effective diffusion coefficient of intercalant in porous media (see, e.g. Refs.8,11).

In fact, the driving force in the process of LIB charging or discharging is the gradient of electrochemical potential of the components, which determines the process of ion transfer in this system, including electrochemical reactions at the interface of the electrode particles. Thus, it is very important to develop new models of ion transport in lithium-ion batteries, which will enable operation directly with the chemical potentials of the components instead of their concentrations analyzed in SP and P2D models. This result is achievable with the phase-field theory,[13-18] which is widely used in the theory of phase transitions.[19]

Application of the phase-field theory to analysis of intercalation and transport of ions in electrochemical systems was launched by Guyer *et al.* who developed the model of electrodeposition of ions at the sharp electrode/electrolyte interface.[16,17] The dynamics of intercalant concentration was described in terms of the Cahn-Hilliard (CH) equation and regular solution approximation for interaction of the solute and solvent. Here, the electrode/electrolyte interface was determined by the order parameter $\xi$ varying from 0 to 1 with the dynamics predicted by the Allen-Cahn equation.

Han *et al.*[18] applied the CH equation to calculate ion diffusion in a single electrode particle in one-dimension. The diffusion was obtained in the form of concentration waves that differed from classical SP and P2D models. The process of insertion/extraction of ions provokes deformation of electrode particles and can lead to the elastic stress of matter. The effect of elastic stress on free energy density and CH dynamics of intercalant was studied by Tang *et al.*[20] and Huttin *et al.*[21]

Fleck *et al.* developed the model of diffusion limited solid-solid phase transformation during lithium insertion in LiFePO$_4$-nano-particles.[22] The model describes propagation of the lithium concentration front in a single particle by the CH equation with anisotropic mobility, whereas transformation of the particle shape is described in terms of the Allen-Cahn equation taking into account elastic energy.

As mentioned above, the BV relation is one of the basic relations that gives correct quantitative and qualitative conclusions in the classical theory of electrochemical systems.[2-12] Therefore, the combined models employing the phase-field theory and BV relation were developed. The first model of this type was developed by Bazant and coauthors[23-26] who introduced exponential Tafel dependence of the current on the overpotential, defined in terms of the variational chemical potentials.[23] Also, this approach was employed for insertion/extraction process, represented as propagating concentration front, for analysis of the effect of particle size[24] and elastic coherency strain[25], and to demonstrate the suppression of the concentration front formation in LiFePO$_4$ nanoparticles with increasing current.[26]

Another attempt to combine phase-field theory and Battler-Volmer relation was made by Zelič and Katrašnik.[27] They modified the CH equation with introduction of the source term related to the Buttler-Volmer relation. Here, the elastic strain was also taken into account.

Liang *et al.*[28,29] developed the model of ion electrodeposition on the diffuse electrode/electrolyte interface. The model accounts the diffusion of ions in electrolyte by the ambipolar diffusion equations with a source term. The process of ion electrodeposition at the electrode interface is described in 1D with the modified Allen-Cahn equation including extraterm formulated in consistency with the BV relation.

Phase-field theory was applied to investigate the formation and growth of a solid electrolyte interface (SEI) layer on the anode surface of a lithium ion battery by Deng *et al.*[30] In this model, ions and electrons are transported by diffusion and electromigration with concentration-dependent mobility. The SEI formation reaction rate is determined by the generalized Butler-Volmer equation.

Phase-field model for chemo-mechanical induced fracture in lithium-ion battery electrode particles was proposed by Miehe *et al.*[31]. They applied a geometric approach to the diffusive crack modeling based on the introduction of a global evolution equation of regularized crack surface, governed by the crack phase field. The gradient-extended Cahn–Hilliard equation for the Li-ions with account for possible phase segregation was utilized. Zuo *et al.*[32] proposed a phase-field model for coupling lithium diffusion and stress evolution with crack propagation.

Summarizing the basic assumptions of the phase-field models,[13-33] one can conclude that most of them are related to insertion or extraction process in an individual particle. The interaction is described in regular solution approximation and electrochemical reaction is formulated in terms of modified BV relation. Actually, the electrode particles have essentially non-spherical shape and are characterized by the size distribution function.[34-38] Particles with different size are charged or discharged during different time intervals.[38] The real propagation of the concentration waves in porous media can cause inhomogeneous insertion or extraction of ions over the particle interface. All these peculiarities are not taken into account in classical (SP, P2D) and existing phase-field models directly. Thus, the microscopic approach allowing simulation of the transport and intercalation in batteries in highly inhomogeneous system taking into account the porosity of electrodes, arbitrary shape of particles and size distribution function is required. Most of the models (see e.g. Refs.15-17, 19-29) operate with 1-D simulation, thus they are not suitable for analysis of the percolative nature of intercalant transport in interconnected electrode particles and electrolyte forming the network of the diffusion channels characterized by the different values of the species mobility. Also, all the phase-field models of ion intercalation and transport are performed in regular solution approximation, thus they are applicable for molecules with short-range interaction and approximately equal size. Nevertheless, the molar volume of intercalant (e.g. lithium) is less than the molar volume of solvent molecules (e.g. olivines, graphite etc.) violating the applicability of regular solution approximation.

In this study, we develop the microscopic 3D phase-field model of intercalation and transport of ions in lithium-ion batteries with realistic distribution of electrode particles taking into account percolative transport of ions in porous electrodes and difference in size of solute and solvent molecules. In this approach, we simulate the battery as a whole including cathode, anode, and separator. The basic feature of this model is in introduction of smooth dependences of transport and interaction parameters over the system volume, where the distribution of electrode material and electrolyte is determined by the space dependence of order parameter. The driving force of the intercalation process is the difference of the electrochemical potentials in electrode particles and electrolyte. The proposed approach allows us to avoid the introduction of the special boundary conditions at the particle interface that differs from the other models employed for modeling of lithium-ion batteries functioning.

## 2. Phase-Field Model of Intercalation in Electrode Particles

Let us consider the intercalation and transport of intercalant (e.g. lithium) in a battery in terms of the phase-field theory. We assume the battery comprising of anode, cathode and separator as in an ordinary construction of a battery.[1-12] The space between electrode particles and separator is filled with electrolyte. We describe this system as inhomogeneous medium with interaction and transport parameters depending on the order parameters $\xi_a$ and $\xi_c$ having value of 0 in electrolyte and 1 in electrode particles. The value varying between 0 and 1 corresponds to electrode particles interface where the electrochemical reaction takes place.

The modified CH equation taking into account the interfacial electrochemical reaction and the spatial dependence of interaction (see e.g. Ref. 39) and dynamic parameters can be generally written in the form

$$\frac{\partial c}{\partial t} = \nabla \left[ M \nabla \left( n_0 \frac{\partial g}{\partial c} - \nabla \kappa \nabla c \right) \right] - \nabla \mathbf{j}_{\text{er}}. \qquad (1)$$

Here $c \equiv c(\mathbf{r},t)$ is the concentration field of intercalant varying between 0 and 1, $M \equiv M(\xi)$ is the mobility, $\kappa \equiv \kappa(\xi)$ is the gradient energy coefficient, $g(\xi,c)$ is the free energy density of the binary alloy, where we use order parameter $\xi = \xi_a(\mathbf{r})$ in anode and $\xi = \xi_c(\mathbf{r})$ in cathode, $n_0$ is the number of atom per unit volume. Also, we include here the flux density $\mathbf{j}_{\text{er}}$ describing the electrochemical reaction at the interface of anode and cathode particles.

Usually, atomic volume of intercalating atom (solute) is less than molecules of the electrode particles (solvent) that can be accounted within the theory of mixtures developed by Guggenheim[40]. The general relation for the free energy density of the solid solution reads

$$g(\xi,c) = \mu_2^0 + \left(\mu_1^0 - \mu_2^0\right)c + c(1-c)\left[\Omega^0 + \Omega^1(1-2c) + \Omega^2(1-2c)^2 + ...\right] + $$
$$k_BT\left[(1-c)\ln\left\{\frac{1-c}{1+c(\rho-1)}\right\} + c\ln\left\{\frac{\rho c}{1+c(\rho-1)}\right\}\right]. \qquad (2)$$

Here, $\mu_1^0$ and $\mu_2^0$ are chemical potentials of pure solute and solvent, respectively, $\Omega^i \equiv \Omega^i(\xi)$ are interaction parameters ($i=0,1,2...$) depending on the ordering, $T$ is temperature, $\rho$ is the ratio of atomic volumes of solute and solvent, $k_B$ is the Boltzmann constant. Here, volume of electrode particles is assumed to be unchangeable during intercalation. Some general aspects of applicability of the used Redlich–Kister power series in Eq.2 for calculation of the phase equilibrium in binary and multicomponent solutions are discussed in Ref. 41.

In this model, we assume that interaction and dynamic parameters depend on the ordering $\xi$ in the form

$$\mu_i^0(\xi_{c,a}) = \mu_{ic,a}^0 h(\xi_{c,a}) + \mu_{ie}^0(1 - h(\xi_{c,a})),$$
$$M(\xi_{c,a}) = M_{c,a} h(\xi_{c,a}) + M_e(1 - h(\xi_{c,a})),$$
$$\Omega^i(\xi_{c,a}) = \Omega_{c,a}^i h(\xi_{c,a}) + \Omega_e^i(1 - h(\xi_{c,a})), \qquad (3)$$
$$\kappa(\xi_{c,a}) = \kappa_{c,a} h(\xi_{c,a}) + \kappa_e(1 - h(\xi_{c,a})),$$

where subscripts "c", "a" and "e" denote the values attributed to cathode, anode, and electrolyte, respectively. The function $h(\xi_{c,a})$ provides smooth change of the parameters in Eq.3 over the interfacial layer between cathode (c) or anode (a) particles and electrolyte. This function should have two extreme points at $\xi=0$ and $\xi=1$. Therefore, the power law family for this function can be written in the form $h(\xi) = A\int \xi^n(1-\xi)^m d\xi$, where $m$ and $n$ are real numbers ($m, n \geq 1$), and $A$ is a normalization constant providing the equality $h(1)=1$. Generally this function can be written with hypergeometric function. In this study, we put $m=n=2$ that gives $h(\xi) = \xi^3(6\xi^2 - 15\xi + 10)$. This function is ordinarily applied to determine spatial dependence of physical quantities in phase-field theory.[19,28,29]

For simplicity, the constant values of the gradient energy coefficient ($\kappa \equiv \kappa_c = \kappa_a = \kappa_e$), and zero values of high order interaction parameters ($\Omega_{c,a,e}^0 > 0$ and $\Omega_{c,a,e}^i = 0$ for all $i > 0$) are employed. Also, we assume the solute mobility to be independent of the solute concentration $c$ ($M_{c,a,e}$ = const) and can be coupled with corresponding value of the diffusion coefficient by the Einstein relation $n_0 M_{c,a,e} = D_{c,a,e}/(k_BT)$. Generally, the mobility of intercalant can depend on the local concentration, e.g. in different phase-field models the linear or degenerative dependence can be employed for the analysis of the new phase formation (see e.g. Refs. 42, 43).

The flux density $\mathbf{j}_{er}$ corresponds to the rate of electrochemical reaction appearing in the interfacial layer of the electrode particles described by the space distribution of ordering in cathode $\xi_c$ and anode $\xi_a$. The interfacial flux density can be obtained by the Butler-Volmer relation

$$\mathbf{j}_{er}(\xi_{a,c}) = j_{BV}\mathbf{n}, \quad j_{BV} = j_0\left[\exp\left(\frac{\alpha z e \eta_{a,c}}{k_BT}\right) - \exp\left(-\frac{(1-\alpha)z e \eta_{a,c}}{k_BT}\right)\right], \qquad (4)$$

where $j_0$ is a constant, $\eta_{a,c}$ is the overpotential, $z$ is the valence of the intercalating atom in ionized state, $e$ is the elementary charge, $\mathbf{n}$ is a normal at the electrode particle interface that can be determined as $\mathbf{n} = \nabla h/|\nabla h| \approx l_0 \nabla h$, $l_0$ is the thickness of interfacial layer. If the overpotential does not change the absolute value of interfacial flux, density is constant $|\mathbf{j}_{er}| = j_{BV}$. Also, the flux density is expected to be proportional to mobility of intercalant $j_{BV} \sim n_0 M(\xi_{a,c})$. Thus, the flux density of the electrochemical reaction can be introduced as

$$\mathbf{j}_{er}(\xi_{a,c}) = -M(\xi_{a,c}) n_0 \Delta\mu_{a,c} \nabla[h(\xi_{a,c})].$$

Here $\Delta\mu_{a,c}$ is a constant associated with a shift of chemical potential due to electrochemical reaction and depending on overpotential. If the overpotential is small ($ze\eta_{a,c} \ll k_B T$) linear expansion of the Eq.4 gives $\Delta\mu_{a,c} \sim ze\eta_{a,c}$.

Summarizing the above assumptions and approximations we can transform Eq.1 to the ordinary form of CH equation

$$\frac{\partial c}{\partial t} = \nabla\left[M(\xi_{a,c})\nabla\left(n_0 \frac{\partial \tilde{g}}{\partial c} - \kappa \nabla^2 c\right)\right].$$

with the effective free energy density in the form of

$$\tilde{g}(\xi_{a,c}, c) = \mu_2^0(\xi_{a,c}) + \left[\mu_1^0(\xi_{a,c}) - \mu_2^0(\xi_{a,c}) + \Delta\mu(\xi_{a,c})\right]c + \Omega^0(\xi_{a,c})c(1-c) +$$

$$k_B T\left[(1-c)\ln\left\{\frac{1-c}{1+c(\rho-1)}\right\} + c\ln\left\{\frac{\rho c}{1+c(\rho-1)}\right\}\right],$$

where $\Delta\mu(\xi_{a,c}) = \Delta\mu_{a,c} h(\xi_{a,c})$. The effective free energy density $\tilde{g}(\xi_{a,c}, c)$ takes into account the change of the chemical potential emerging due to electrochemical reaction at the electrode particle interface.

For further analysis of the intercalation and transport the following dimensionless parameters have been introduced

$$\mathbf{r}^* = \frac{\mathbf{r}}{l}, t^* = t\frac{n_0 M_e \Omega_c}{l^2}, l^2 = \frac{\kappa}{n_0 \Omega_c}, T^* = \frac{T}{T_C}, T_C = \frac{\Omega_c}{2k_B},$$

$$g^* = \frac{\tilde{g}}{\Omega_c^0}, \Omega_{a,c,e}^* = \frac{\Omega_{a,c,e}^0}{\Omega_c^0}, M_{a,c,e}^* = \frac{M_{a,c,e}}{M_e}. \tag{5}$$

Here, $l$ is the correlation length in cathode that is accepted as a scale in space. Observation of gradual transport of intercalant between the electrodes through the electrolyte requires $M_a^* \ll M_e^*$ and $M_c^* \ll M_e^*$ ($M_e^* = 1$). Otherwise, we can observe fast extraction of ions from electrode particles and slow transport in electrolyte since the size of particles is much less than a linear size of the whole battery. Also, fast extraction and relatively slow diffusion in electrolyte can be responsible for a local increase in concentration between particles of porous electrode. High concentration can cause formation of stable nuclei in electrolyte phase if the corresponding interaction parameter and the temperature allow their formation.

Taking into account the molar concentration of electrode materials[4-6] and pure lithium (76340mole/m$^3$) we can calculate the ratio of atomic volumes for different electrode materials as $\rho = 0.30$ (Li$_c$FePO$_4$, 22900 mole/m$^3$)[27], $\rho = 0.31$ (Li$_c$Mn$_2$O$_4$, 23720mole/m$^3$)[6], $\rho = 0.35$ (Li$_c$C$_6$, 26400mole/m$^3$)[6], $\rho = 0.38$ (Li$_c$TiS$_2$, 29000mole/m$^3$)[5]. In this study, we assume that the ratio has a constant value $\rho = 0.3$ in both electrodes and electrolyte. The value of $n_0$ is also assumed to be constant over all system.

The phase-field theory assumes that formation of the phase enriched in solute can occur in any place of the system if the nucleation barrier is overcome.[44-46] If the initial state of the system corresponds to domain of unstable states, the phase formation is barrierless and occurs by the spinodal decomposition mechanism.[44-47] Generally, it can be also related to electrolyte phase. Formation of nuclei in electrolyte can stop the transport between electrodes and cause degradation of the battery. Thus, the temperature of the battery functioning should be higher than critical temperature of electrolyte and lower than that of the electrode. In this study, we use the following values of dimensionless interaction parameters $\Omega_c^* = \Omega_a^* = 1$ and $\Omega_e^* = 0.5$. The negative formation of stable nuclei in electrolyte has been observed by the example of the battery with interaction parameters $\Omega_c^* = \Omega_a^* = 1$ and $\Omega_e^* = 1.1$.

The introduced dimensionless interaction parameter in electrodes is coupled with the dimensionless temperature. The interaction parameter for lithium in different cathode materials in regular solution approximation is estimated in the range 0.059-0.189eV.[18,20-27] If we assume that charging or discharging of the battery occurs at the temperature of 300K, the plausible dimensionless temperature can lie in the wide range of $0.27 < T^* < 0.87$. In this study, we employ the value of $T^* = 0.56$.

The phase diagram for electrode material can be calculated by the theory of phase equilibria taking into account the difference of atomic volumes of solute and solvent. The equilibrium solute concentration $c^\alpha$ and $c^\beta$ in $\alpha$ and $\beta$ phase should satisfy the fundamental relations[48]

$$\frac{\partial g^{*\alpha}}{\partial c^\alpha} = \frac{\partial g^{*\beta}}{\partial c^\beta}, \qquad g^{*\alpha} - c^\alpha \frac{\partial g^{*\alpha}}{\partial c^\alpha} = g^{*\beta} - c^\beta \frac{\partial g^{*\beta}}{\partial c^\beta}, \qquad (6)$$

Here, we use the modified free energy density $g^*(c,\xi)$ taking into account the shift of the chemical potential due to interfacial electrochemical reaction. Appling the introduced interaction parameters, we calculate the phase diagram for electrode material that is shown in Fig.1. The points corresponding to equilibrium concentration, metastability limit and critical point for the considered temperature are also marked in this figure. The critical temperature of the considered solution is $T^* = 0.79$ and it differs from the value of $T^* = T_C^* = 1$ calculated in regular solution approximation ($\rho = 0$, $\Omega_{c,a,e}^i = 0$ for all $i > 0$).

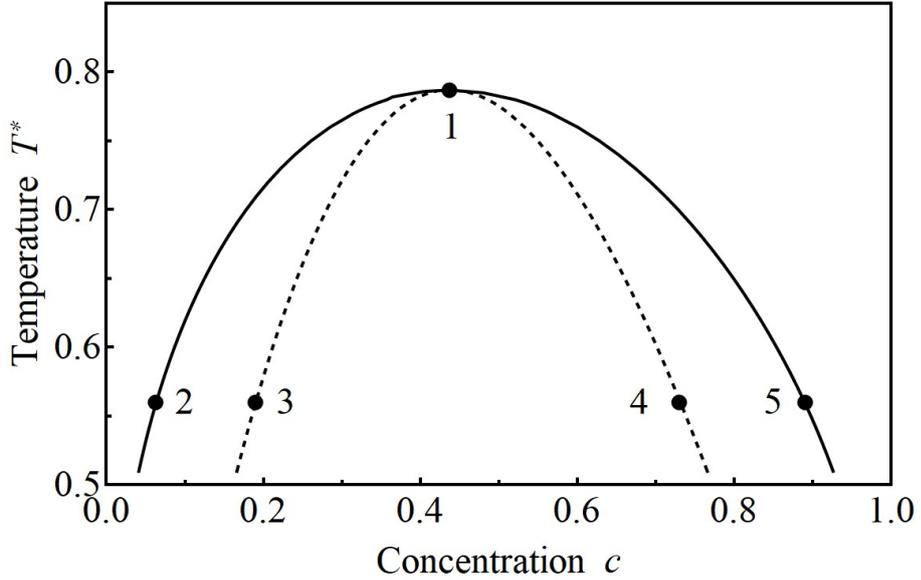

**Figure 1.** Phase diagram of the binary alloy described by the free energy density (2). Solid line is the concentration of equilibrium phases, dashed line is the metastability limit (spinodal line). The critical point 1 is characterized by the concentration of 0.44 and temperature $T^* = 0.79$. Equilibrium concentration (points 2 and 5) has the values of 0.063 and 0.89. Metastability limit (points 3 and 4) has the values of 0.19 and 0.73.

The scenario of the system evolution can be predicted by the analysis of the extended free energy density landscape $g^*(c,\xi)$ shown in Fig. 2. The figure consists of two conjugated parts and unifies data on anode ($0 \le \xi_a \le 1$) and cathode ($0 \le \xi_c \le 1$) with different values of $\Delta\tilde{\mu}$, which are $\Delta\tilde{\mu}_a$ and $\Delta\tilde{\mu}_c$, respectively. The example demonstrates that change of these two values can cause formation of preferable state characterized by the lower value of free energy density. Thus, we expect that Fig. 2(a) describes the system with flow of intercalant from cathode to anode, and Fig. 2(b) describes the flow in the backward direction. The process is accompanied with the overcoming of the barrier (Fig. 2) that takes place due to interfacial electrochemical reaction defined by values of $\Delta\tilde{\mu}_a$ and $\Delta\tilde{\mu}_c$.

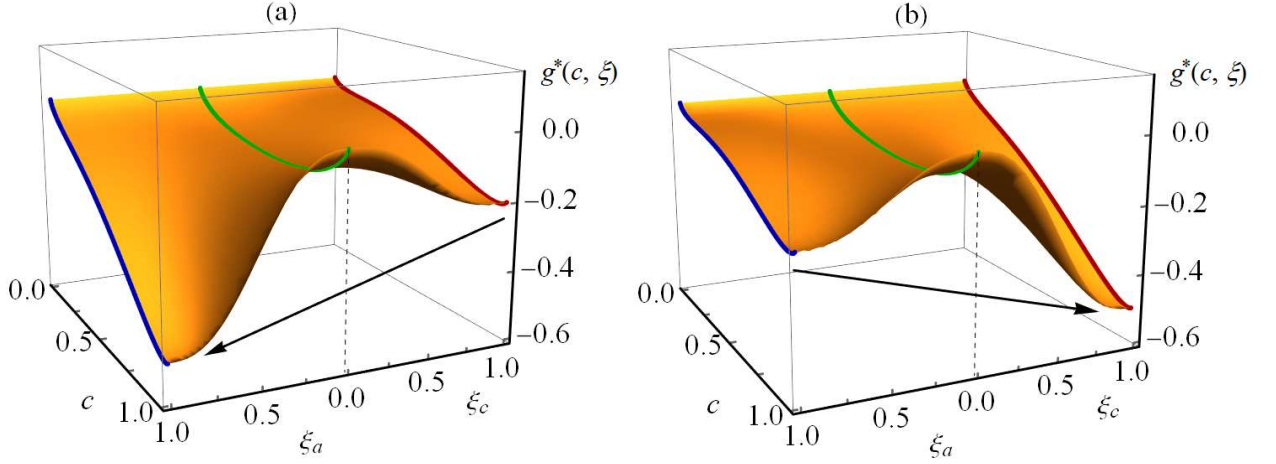

**Figure 2.** The extended free energy landscape for anode, cathode and electrolyte material. Thick lines for $\xi_a = 1$, $\xi_c = 1$ and $\xi_a = \xi_c = 0$ correspond to pure anode, cathode and electrolyte materials. Vertical dashed line divides the diagram into two parts describing the free energy at the anode-electrolyte and cathode-electrolyte interface. Figure (a) is the landscape calculated for $\Delta\tilde{\mu}_a = -0.5$ and $\Delta\tilde{\mu}_c = -0.2$. Figure (b) is calculated for $\Delta\tilde{\mu}_a = -0.2$ and $\Delta\tilde{\mu}_c = -0.5$. The arrows schematically show the direction of transition to the state with lower free energy density.

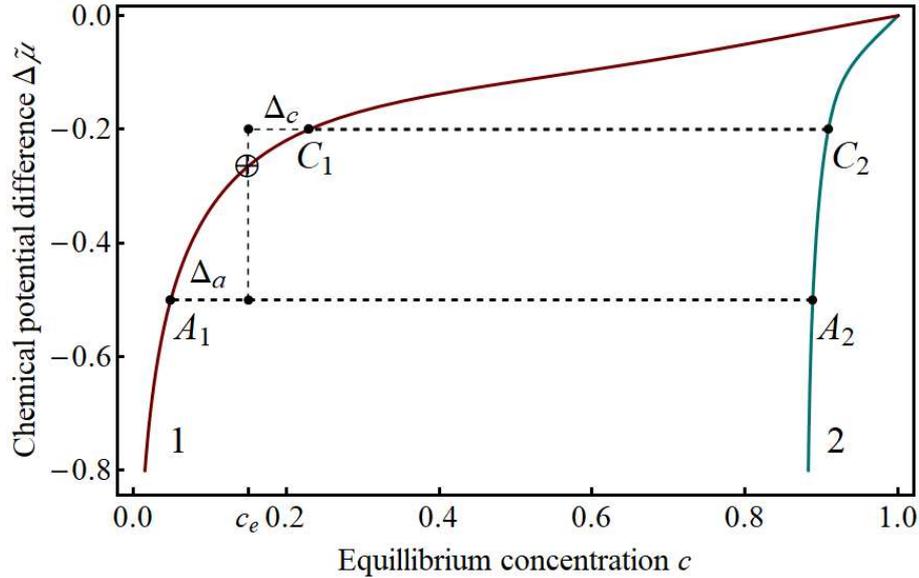

**Figure 3.** The phase diagram of electrode-electrolyte equilibrium at the temperature $T^* = 0.56$. Points C1 and C2 show the equilibrium concentrations in cathode for $\Delta\tilde{\mu}_c = -0.2$. Points A1 and A2 show the equilibrium concentration in anode for $\Delta\tilde{\mu}_a = -0.5$. Lines show the dependence of equilibrium concentration on chemical potential difference in electrolyte (1) and electrode (2). Circle with a cross shows the equilibration point for electrolyte concentration $c_e$, $\Delta_a$ and $\Delta_c$ shows excess and lack of solute in electrolyte for anode (A1-A2) and cathode (C1-C2).

This conclusion can be also supported by the analysis of electrode-electrolyte phase diagram (Fig. 3) that is calculated by the general thermodynamic relations Eq.6. If the concentration $c_e$ is between A1 and C1, we have excess of solute in electrolyte for anode-electrolyte equilibrium ($\Delta_a > 0$). Hence, the anode should absorb the solute from electrolyte. The concentration of solute lacks in cathode $(\Delta_c < 0)$. Therefore, the extraction of solute from cathode particles should be observed. If concentration in electrolyte coincides with A1 or C1, the system of electrode-electrolyte is in equilibrium and flux of the solute through the electrode particles interface is absent. In practice, the intercalation process can also depend on other factors, e.g. state of the electrode particle and electrolyte, heterogeneity of intercalation process, influence of the diffuse interface etc. Thus, the equilibrium point can change dynamically during charging or discharging of the battery.

The dimensionless mobilities in cathode $M_c^*$ and anode $M_a^*$ can be estimated by the experimental data on diffusion coefficient. Some of the experimental data on lithium diffusion coefficient for different materials of electrode and electrolyte are summarized in Table 1. The values of the dimensionless mobilities in cathode and anode particles can vary in the range $10^{-6} < M_{c,a}^* < 10^{-1}$. Also, the diffusion coefficient in anode has a higher value than in cathode (see Table 1), therefore the applied values of the dimensionless mobilities should satisfy the inequality $M_c^* < M_a^*$. In this study, the application of the developed model is demonstrated with the values of the dimensionless mobilities $M_c^* = 0.05$ and $M_a^* = 0.075$. Generally, the diffusion coefficient of lithium in electrode material and electrolyte depends on its concentration (see e.g. Refs. 5,6,10,51). This chemical mobility can be easily introduced into numeric solution of equation Eq.1, but it has higher computational cost related to lower values of step in space and time necessary to provide numerical scheme stability and accuracy. Therefore, in this work we employ the intercalant mobility depending on the order parameter only.

**Table 1.** Data on the lithium diffusion coefficient in different materials of electrolyte, cathode and anode employed in simulation of lithium-ion batteries

| $D_c$, m$^2$/s | $D_a$, m$^2$/s | $D_e$, m$^2$/s | Ref. |
|---|---|---|---|
| $5 \cdot 10^{-13}$ (TiS$_2$) | - | $7.5 \cdot 10^{-12}$ (PEO-LiCF$_3$SO$_3$) | Ref.5 |
| $10^{-13}$ (Li$_x$Mn$_2$O$_4$) | $5 \cdot 10^{-13}$ (Li$_x$C$_6$) | $2.58 \cdot 10^{-10}$ (propylene carbonate) | Ref.6 |
| $2.0 \cdot 10^{-16}$ | $3.7 \cdot 10^{-16}$ | $2.6 \cdot 10^{-10}$ | Ref.49 |
| $1 \cdot 10^{-14}$ | $3.9 \cdot 10^{-14}$ | $7.5 \cdot 10^{-10}$ | Ref.50 |

Thus, the dimensionless CH equation describing intercalation and transport in the modeled battery reads as

$$\frac{\partial c}{\partial t^*} = \nabla^* \left[ M^* \nabla^* \left\{ \frac{\partial g^*}{\partial c} - \kappa^* \nabla^{*2} c \right\} \right], \quad (7)$$

where the derivative of dimensionless free energy density is determined in the form

$$\frac{\partial g^*}{\partial c} = \Delta \tilde{\mu}(\xi) + \Omega^*(\xi)(1-2c) + \frac{1}{2} T^* \left[ \ln\left(\frac{c\rho}{1-c}\right) - \frac{\rho-1}{1+c(\rho-1)} \right],$$

and $\Delta \tilde{\mu}$ is the dimensionless difference of chemical potentials $\Delta \tilde{\mu} = (\mu_1^0 - \mu_2^0 + \Delta \mu) / \Omega_c^0$. All dimensionless dynamic and interaction parameters in Eq.7 and the chemical potential difference $\Delta \tilde{\mu}$ have the space dependence of magnitude through the order parameter in the form of Eq.3.

The numeric solution of the dimensionless CH equation Eq.7 is performed by the explicit finite-difference method. The numeric scheme was implemented as GPU–based program, with employment of Titan V videocard for simulation. The time step is $\Delta t^* = 5 \cdot 10^{-3}$.

## 3. The Influence of Intercalant Concentration on Insertion and Extraction process

Let us discuss the influence of intercalant concentration on insertion and extraction by the example of an individual spherical particle. The particles are embedded into the uniform electrolyte and exhibits formation of the spherical concentration front propagating from interface to center of the particle during insertion or extraction process (see Fig. S2, S3 and Ref. 39).

We simulate the intercalation in spherical particles for different values of chemical potential difference, initial concentration in particles and electrolyte. The dimensionless average flux density

$$j_n^* = -\frac{1}{4\pi R^{*2}} \frac{d}{dt^*}\left[\sum_{\xi_i \geq \xi_{th}} c_i\right], \qquad (8)$$

is employed for analysis of the process. Summation in Eq.8 is performed over the all particle nodes with ordering exceeding the threshold value $\xi_{th} = 0.9$. The dimensionless flux density can be easily coupled with the current density $j_n^e$ by the relation $j_n^* = j_n^e \tau l^2 /(ze)$, where we use the coefficient $\tau = l^2 T^* /(2D_e)$.

The flux density $j_n^*$ is calculated at the beginning of the insertion or extraction when concentration of intercalant in particles and electrolyte does not change significantly. Fig. 4 shows the dependence of the flux density $j_n^*$ on chemical potential difference in an individual electrode particle for the different values of initial concentration in particle and electrolyte. The value of the chemical potential difference $\Delta\tilde{\mu}_0$ corresponding to a change of the flux density sign also depends on the concentration in electrode particles and electrolyte (see also Fig. 3). This influence is explained by change of the solute chemical potential in electrolyte or/and particles with a change of concentration therein. Thus, we expect that higher concentration in electrolyte can stop the extraction from the enriched electrode particles. Furthermore, we conclude that inhomogeneous distribution of intercalant concentration in porous electrodes at the constant value of overpotential can cause the inhomogeneous distribution of current density.

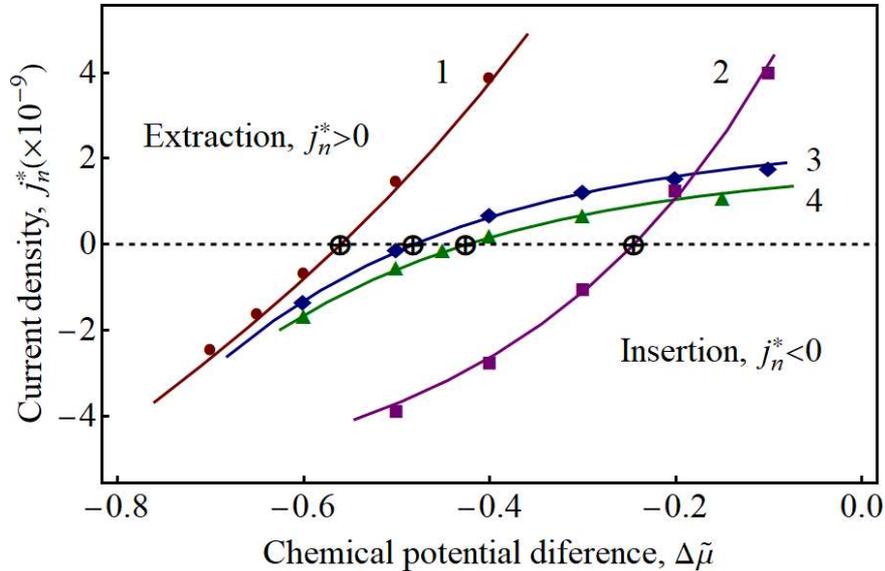

**Figure 4.** The dependence of interfacial flux density on the chemical potential difference calculated by direct solution of the CH equation for spherical cathode particles ($R_c = 10$) with different values of initial concentration in electrolyte, electrode particle: circles — $c_e = 0.07, c_c = 0.89$, diamonds — $c_e = 0.07, c_c = 0.1$ triangles — $c_e = c_c = 0.07$, squares — $c_e = 0.2, c_c = 0.9$. The temperature is $T^* = 0.56$. Solid lines are approximations to guide the eye. Black circles with a cross show the point of equilibration of electrode and electrolyte, where the interfacial current density equals zero ($\Delta\tilde{\mu}_{01} = -0.56, \Delta\tilde{\mu}_{03} = -0.48, \Delta\tilde{\mu}_{04} = -0.43,$ $\Delta\tilde{\mu}_{02} = -0.25$).

## 4. Space Dependence of Ordering in Porous Electrodes

To construct a three-dimensional model of the electrodes, we chose the approach proposed by Feinauer *et al.* in paper.[36] The authors use random Gaussian fields on a sphere as models of particle distribution in porous electrode. This mathematical model, validated by the experimental data,[36] can reproduce realistic microstructures of lithium-ion battery anodes, which can be used in other models related for simulation of percolative transport of intercalant. A detailed description of the model and simulation algorithm can be found in Refs. 36 and 37. We construct the electrodes taking into account the boundary conditions employed for the solution of the CH equation (Eq.7). In contrast to the original approach, the boundary conditions along the *y* and *z* axes are periodic. The free boundary conditions are used along the *x* axis. Importantly, the neighbor particles can intersect with each other enabling exchange of solute between them. The algorithm of the porous electrode formation is described in Supplementary material.

Calculation of the electrode particle distribution allows us to obtain the space dependence of the order parameter $\xi(\mathbf{r})$ approximated by the hyperbolic tangent at the particle interface. To prevent a drastic change of the order parameter that can decrease the numeric scheme stability, we smooth its dependence in space.

This approach of electrode formation is applied for construction of both anode and cathode. Generally, distribution of matter in the modeled electrodes can be performed optionally taking into account an arbitrary construction of the battery.

## 5. Simulation Results and Discussion

Let us apply the developed model to simulate the process of intercalation and transport in the battery consisting of the anode, cathode and separator.

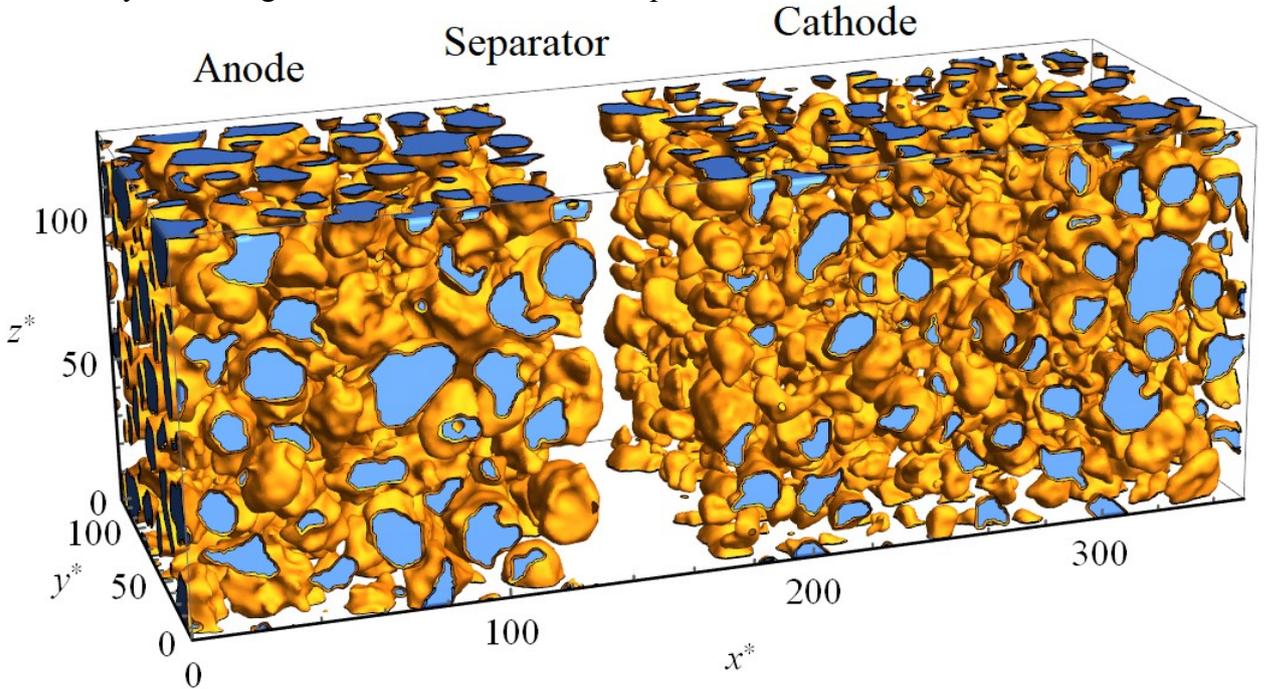

**Figure 5.** The distribution of electrode particles in the simulated battery shown by the iso-surfaces of ordering ($\xi_{a,c} = 0.5$).

The anode and cathode particles fill the parts of the system volume for the coordinate range $0 \leq x^* < X_a$ and $X_c \leq x^* < X_{max}$, respectively. The simulation supercell has the size of $350 \times 128 \times 128$. The electrode boundaries $X_a, X_c$ and $X_{max}$ have the values of $X_a = 135, X_c = 165$ and $X_{max} = 350$. The distribution of particles in anode and cathode is performed by the approach described in the previous section. The calculated space dependence of ordering is shown in Fig. 5. The average radius and standard deviation of the particles in cathode and anode are $\langle R_c \rangle = 8.2, \langle R_a \rangle = 11.3$ and $\sigma_c = 1.8, \sigma_a = 2.3$, respectively. The total number of particles is 304 for anode and 766 for cathode. The size distribution functions for cathode and anode particles are shown in Fig. S4. The thickness of interfacial layer is $l_0 \sim 6-7l$.

Fig. 6 shows the dynamics of concentration field during discharging process for constant chemical potential difference in cathode and anode, which are $\Delta\tilde{\mu}_c = -0.2$ and $\Delta\tilde{\mu}_a = -0.5$, respectively. Initially, the cathode particles have the concentration of 0.89, anode particles and electrolyte have the concentration of 0.068. The 2D and 3D presentation of the intercalation dynamics can be found in Supplementary Materials. At the very beginning of the discharging process, a portion of solute is extracted from cathode particles to electrolyte (Fig. 6a). At this stage, the smallest cathode particles loose all solute and are excluded from the further extraction. The concentration in electrolyte increases that causes the decrease of extraction rate from large particles (see also Fig. 4). Also, some amount of solute is absorbed from the electrolyte by the anode particles (Fig. 6a and b). Then, the solute extracted from cathode particles gradually diffuses in electrolyte through the separator and achieves anode particles which start to absorb the solute from electrolyte (Fig. 6b). The anode particles closer to separator are filled with the solute the first and stop absorbing of the solute from electrolyte. Penetration of solute between anode particles causes intercalation in the next row of particles. Thus, we can observe the motion of the boundary between charged and uncharged particles in anode that can be considered as a concentration front propagating in anode. A similar concentration front in cathode is more diffuse and not clearly marked at the same time points in Fig. 6 b and c. At the end of simulation all cathode particles are depleted and all anode particles are filled with the solute (Fig. 6 d). The excess of solute that could not be absorbed by the anode particles remains in electrolyte. It is important that insertion of solute through the electrode particle interface occurs inhomogeneously (see Fig. 6b). Formation of the nuclei starts at the particle interface that underlies the heterogeneous mechanism of the new phase formation in electrode particles. This feature shows the fundamental difference from the classical SP and P2D models implying homogeneous insertion and extraction of solute in electrode particles that can be observed in case of uniform distribution of solute concentration in electrolyte (compare with Fig. S2 and S3). The heterogeneity of the new phase formation can be explained by the inhomogeneous distribution of solute in electrolyte due to electrode porosity. Another reason can be related to a lower value of the nucleation barrier that usually accompanies the process of heterogeneous nucleation.[52,53] Also, all these observations and conclusions can be confirmed for the larger simulation cell ($625 \times 128 \times 128$) having approximately twice larger size of electrodes (see Supplementary, Fig. S5 and S6).

Fig. 7 shows the process of charging that can be observed with permutation of the chemical potential difference ($\Delta\tilde{\mu}_c = -0.2$, $\Delta\tilde{\mu}_a = -0.5$). The initial distribution of solute corresponds to Fig. 6d to complete the charging-discharging cycle. At the very beginning, the cathode particles absorb the solute excess from electrolyte. After this fast stage, all cathode particles become slightly enriched in solute. Further dynamics of charging process including insertion, extraction and transport of solute is similar to the process of discharging (Fig. 6a-d). The duration of charging ($\Delta t^* \sim 50000$) is slightly shorter than discharging interval ($\Delta t^* \sim 60000$) of the modeled battery that is explained by the different initial distribution of intercalant in the battery.

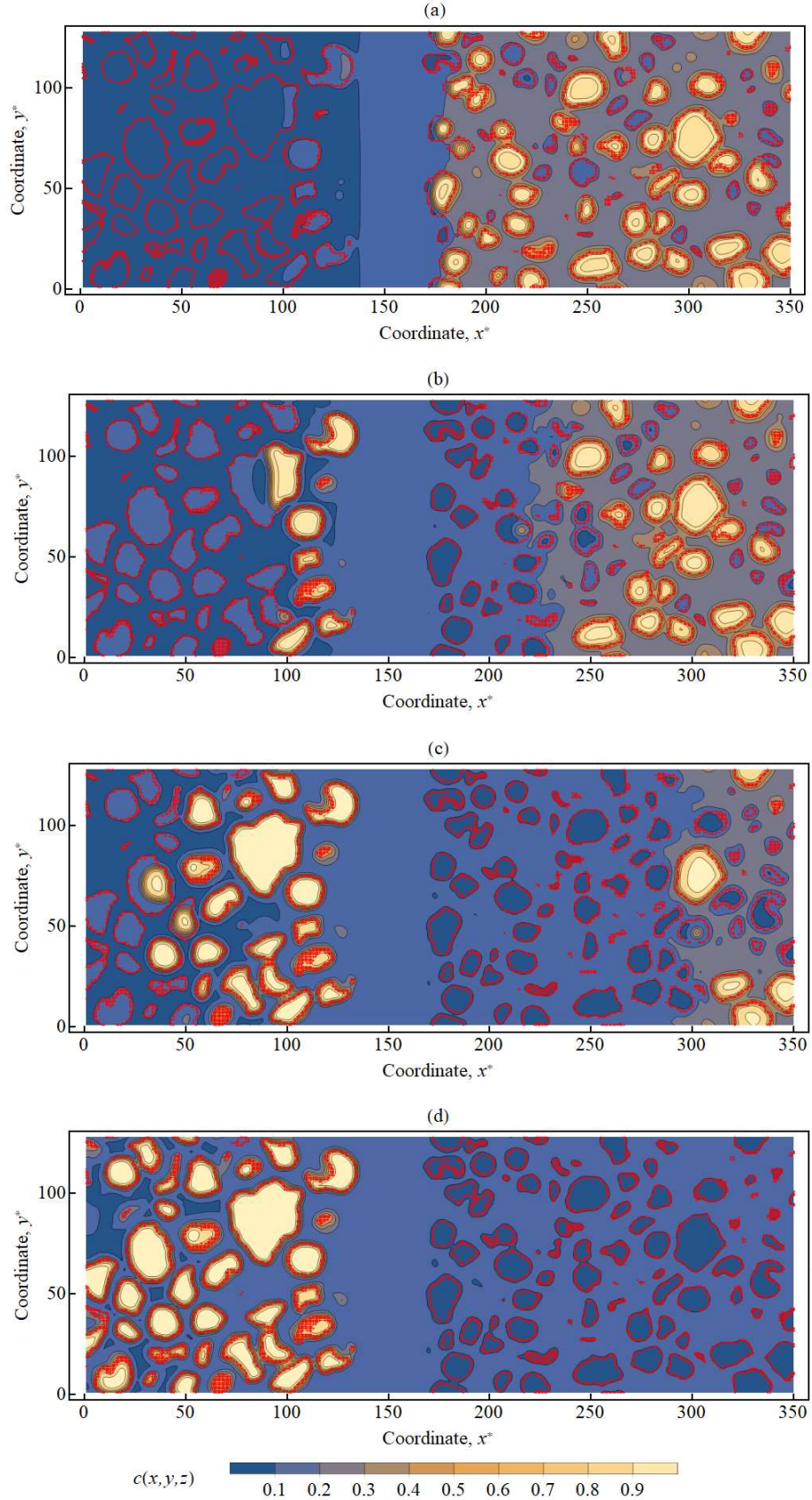

**Figure 6.** Simulation of the dynamics of battery discharging at the constant chemical potential difference in cathode ($\Delta\tilde{\mu}_c = -0.5$) and anode particles ($\Delta\tilde{\mu}_a = -0.2$). Red contours show the particle interface corresponding to the order parameter interval of ($0.5 < \xi < 0.6$). Figures describe different time points: (a) $t^* = 500$, (b) — $t^* = 8000$, (c) — $t^* = 33500$, (d) — $t^* = 60000$. Distribution of concentration is shown in the plane with coordinates $z^* = 32$.

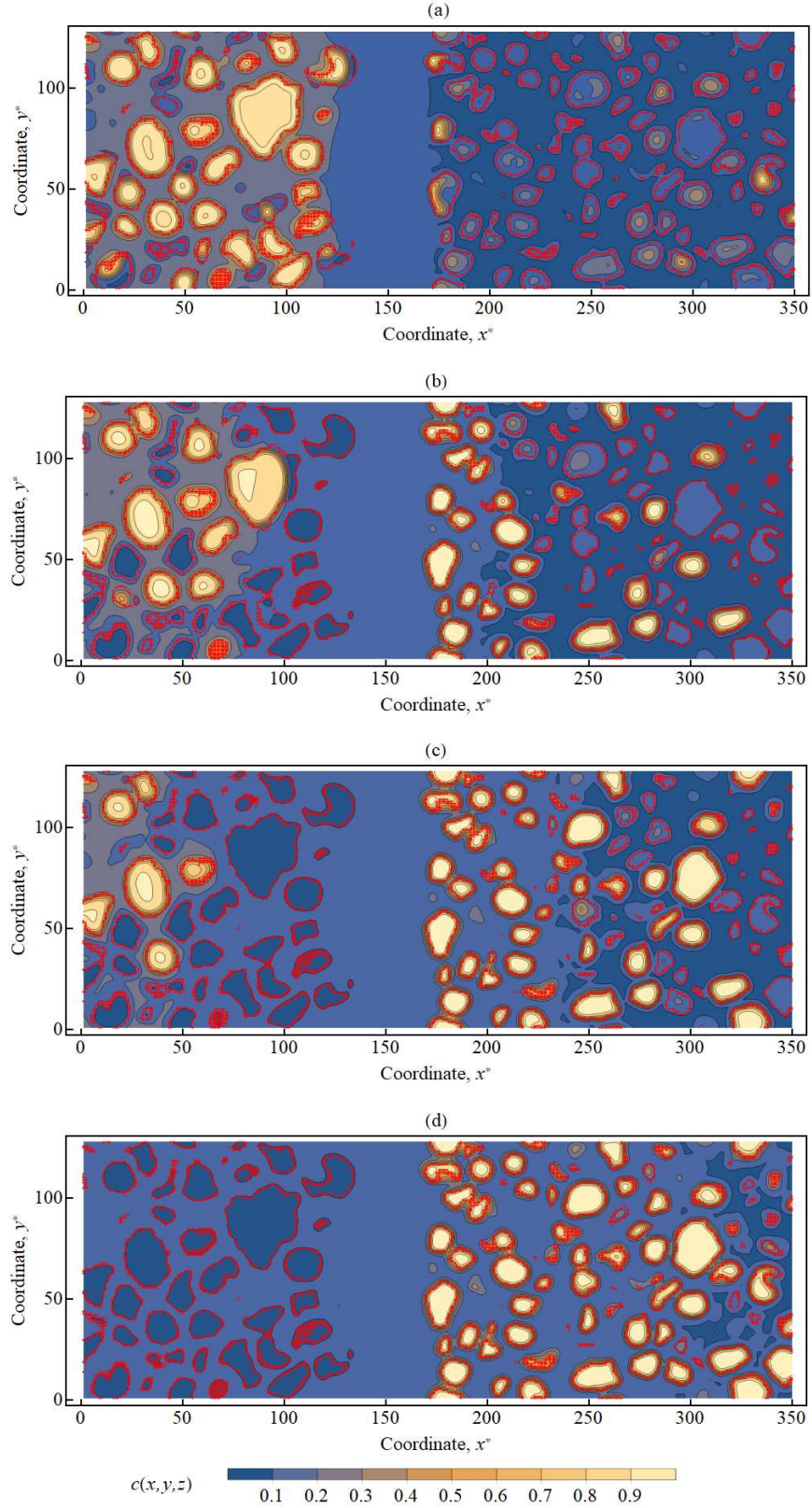

**Figure 7.** Simulation of the dynamics of battery charging at constant chemical potential difference in cathode ($\Delta\widetilde{\mu}_c = -0.2$) and anode ($\Delta\widetilde{\mu}_a = -0.5$) particles. Red contours show the particle interface corresponding to the order parameter interval of ($0.5 < \xi < 0.6$). Figures correspond to different time points: (a) $t^* = 500$, (b) — $t^* = 9500$, (c) — $t^* = 22500$, (d) — $t^* = 50000$. Distribution of concentration is shown in the plane with coordinates $z^* = 32$.

Fig. 8 and 9 show the concentration profiles in electrolyte and electrodes during discharging process. The concentration profiles are obtained by averaging concentration field over coordinates $y^*$ and $z^*$ assuming threshold value of the ordering $\xi_{th} < 0.1$ for electrolyte and $\xi_{th} > 0.9$ for electrodes. Fig. 8 indicates that concentration field in electrolyte increases very fast at the very beginning of discharging process (lines 1 and 2). Then, concentration decreases in cathode and increases in anode gradually (lines 2-6). At the end of simulation the concentration in electrolyte is slightly higher in electrode with lower value of chemical potential difference $\Delta\tilde{\mu}$ that is explained by the enrichment of electrolyte around the particles due to gradual change of overpotential through the particles interface. Dynamics of concentration profile in electrode material (Fig. 9) can be characterized by formation of concentration front that propagates in electrodes from the boundary between the electrodes and separator to the opposite edges of the battery where the current collectors are usually located.

The current density through the anode or cathode current collector ($j_{cc}^*$) can be coupled with the total amount of intercalant absorbed by the electrode particles (see Eq.8). The value of $j_{cc}^*$ can be calculated in the form:

$$j_{cc}^* = \frac{1}{S_{cc}} \left| \frac{d}{dt^*} \left[ \sum_{\xi_i > \xi_{th}} c_i \right] \right|, \qquad (9)$$

where the threshold value of ordering is taken as $\xi_{th} = 0.9$ and $S_{cc}$ is the interface area of electrode current collector ($S_{cc} = 128 \times 128$). Summation in Eq. 9 is performed over all grid points in corresponding electrode, where the ordering $\xi_i$ is higher than an introduced threshold value $\xi_{th}$. The time dependence of the current density can be easily calculated by dynamics of the concentration field (Fig. 6, 7, S5, S6). Calculating the derivative in Eq. 9 with three point difference scheme, we obtain the dynamics of anode current density during the discharging process (Fig. 10). In this figure, the lines show the discharging dynamics for different values of the chemical potential difference in cathode ($\tilde{\mu}_c$) and anode ($\Delta\tilde{\mu}_a$). All the lines exhibit similar discharging dynamics. At the very beginning the current density has the highest value that is related to short-range exchange between electrode particles and electrolyte. The current density decreases and achieves the stage of significantly slower discharging that is coupled with the long-range transport of solute between electrodes and can be related to operation regime of the battery. After this stage, the current density vanishes fast due to completion of discharging process when anode particles are filled with intercalant and cathode particles are completely depleted.

It is also important to compare the discharging dynamics for simulation cells with the different size of electrodes. Lines 3 and 5 in Fig. 10 show the discharging dynamics of two supercells with different size ($350 \times 128 \times 128$ and $625 \times 128 \times 128$). The size of the electrodes is about twice different. The concentration field dynamics corresponding to these lines can be found in Fig. 6, S4, S5. The larger supercell exhibits close value of the anode current density and approximately two times longer discharging time. We have found that charging or discharging process can be expressed as the process of concentration wave propagation through the electrode particles. Therefore, in this model, the result of current density calculation has weak dependence on the electrode size and is determined by the porosity, interaction and transport properties of electrodes and electrolyte. It is expected that increase of the size of electrodes in $x$- direction can cause proportional increase of discharge duration for the fixed values of the chemical potential difference at the electrode interface.

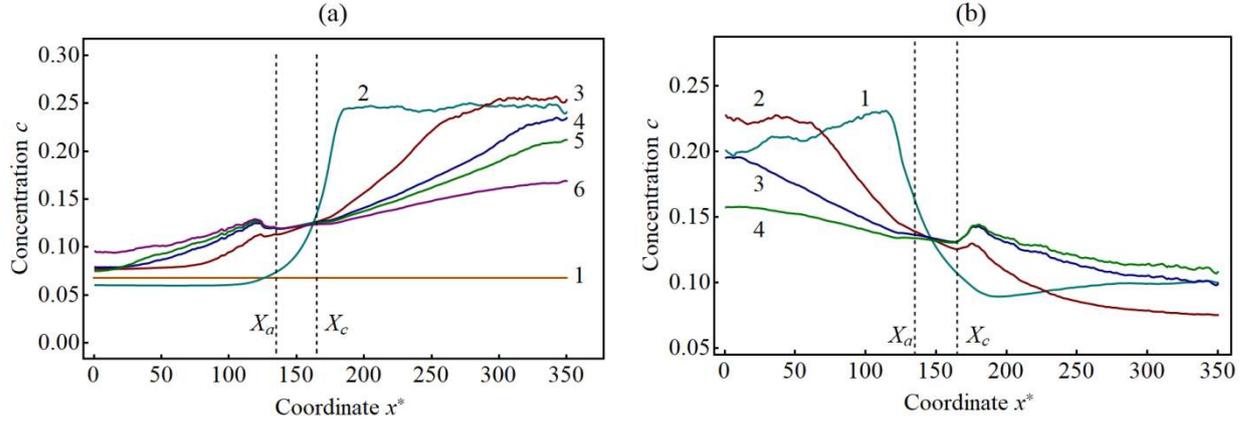

**Figure 8.** Concentration profile of the intercalant in electrolyte during the discharging (figure a) and discharging (figure b) with the constant values of chemical potential differences. Lines in figure (a) correspond to the different time points 1 — $t^* = 0$, 2 — $t^* = 500$, 3 — $t^* = 10000$, 4— $t^* = 30000$, 5 —$t^* = 50000$, 6— $t^* = 60000$. Lines in figure (b) correspond to the different time points: 1 — $t^* = 500$, 2 — $t^* = 10000$, 3—$t^* = 30000$, 4— $t^* = 50000$, 5 — $t^* = 60000$. The chemical potential differences are $\Delta\tilde{\mu}_c = -0.2$, $\Delta\tilde{\mu}_a = -0.5$ (figure a) and $\Delta\tilde{\mu}_c = -0.5$, $\Delta\tilde{\mu}_a = -0.2$ (figure b).

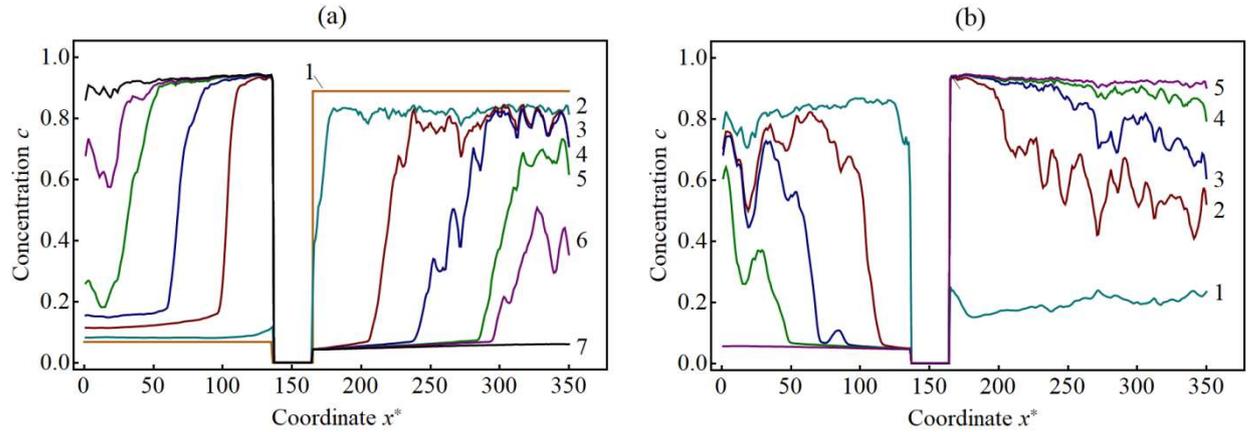

**Figure 9.** Concentration profile of the intercalant in electrodes during the battery discharging (Fig. 6) with the chemical potential differences. Lines in fig. a describe the different time points 1 — $t^* = 0$, 2 — $t^* = 500$, 3 — $t^* = 10000$, 4 — $t^* = 20000$, 5 — $t^* = 30000$, 6 — $t^* = 50000$, 7 — $t^* = 60000$. Lines in fig. b correspond to the different time points 1 — $t^* = 500$, 2 — $t^* = 10000$, 3 — $t^* = 30000$, 4 — $t^* = 50000$, 5 — $t^* = 60000$. The chemical potential differences are $\Delta\tilde{\mu}_c = -0.2$, $\Delta\tilde{\mu}_a = -0.5$ (fig. a) and $\Delta\tilde{\mu}_c = -0.5$, $\Delta\tilde{\mu}_a = -0.2$ (fig. b).

It is interesting to analyze the correlation between the current density at the current collector $j_{cc}^*$ and the chemical potential difference $\Delta\tilde{\mu} = \Delta\tilde{\mu}_c - \Delta\tilde{\mu}_a$. We calculate the average values of anode current density at the stage of slow discharging and corresponding values of the chemical potential difference (see Fig. 11). The dependence satisfies the linear dependence that can be considered as an analogue to the Ohm's law.

It is important to compare qualitatively the developed model with the real systems. The diffusion coefficient and gradient energy coefficient in electrolyte are assumed to be $D_e \sim 10^{-12} \text{m}^2/\text{s}$ and $\kappa \sim 3.13 \cdot 10^9 \text{eV/m}$, respectively.[25] If the molar concentration is $C_m \sim 22900 \text{mole/m}^3$ (Ref. 27) one can easily obtain the value of $n_0 = N_A C_m$ and derive the space and time scaling for the system from the substitution rules Eq.5: $l = [\kappa/(n_0 \Omega_c)]^{1/2} \sim 1.6 \text{nm}$, $\tau = l^2 T^*/(2D_e) \sim 6.9 \cdot 10^{-7} \text{s}$. So, the linear size of the simulation supercell shown in Fig.5 is $L_S = X_{max} l \sim 560 \text{nm}$ and the average particle radius in cathode and anode are $R_c \sim 13 \text{nm}$ and $R_a \sim 18 \text{nm}$, respectively. The interface area of the current collector is $S_{cc} = L_y^* L_z^* l^2 \sim 0.0404 \mu\text{m}^2$. Thus, the considered electrodes consist of nanostructured (0D) materials, which are the most promising for batteries with a high capacity and good rate capability.[54,55] Determination of the space and time scaling of the modeled battery allows us to calculate the current density at operational regime as $j_{cc} = z e j_{cc}^*/(\tau l^2) \sim 93490 \text{A/m}^2 j_{cc}^*$. This calculation gives realistic values of the current density $j_{cc} \sim 5-35 \text{A/m}^2$ (see Figs. 10 and 11) confirming the applicability of the developed approach to calculation of charging and discharging LIB characteristics.

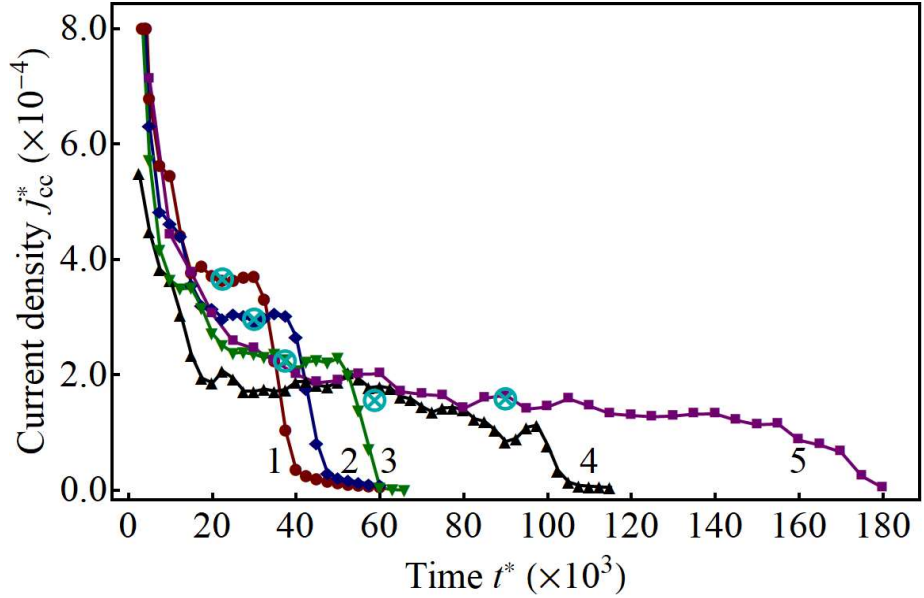

**Figure 10.** Dynamics of the supercell discharging for the different values of chemical potential difference in electrodes: 1 — $\Delta \tilde{\mu}_c = -0.2$, $\Delta \tilde{\mu}_a = -0.7$, 2 — $\Delta \tilde{\mu}_c = -0.2$, $\Delta \tilde{\mu}_a = -0.6$, 3 — $\Delta \tilde{\mu}_c = -0.2$, $\Delta \tilde{\mu}_a = -0.5$, 4 — $\Delta \tilde{\mu}_c = -0.3$, $\Delta \tilde{\mu}_a = -0.5$, 5 — $\Delta \tilde{\mu}_c = -0.2$, $\Delta \tilde{\mu}_a = -0.5$. Crossed circles show the average values of current density corresponding to the stage of slow discharging. Lines 1-4 are calculated for the supercell shown in Fig. 5. Line 5 is calculated for the supercell with a size of 625×128×128 (see Fig. S4 and S5).

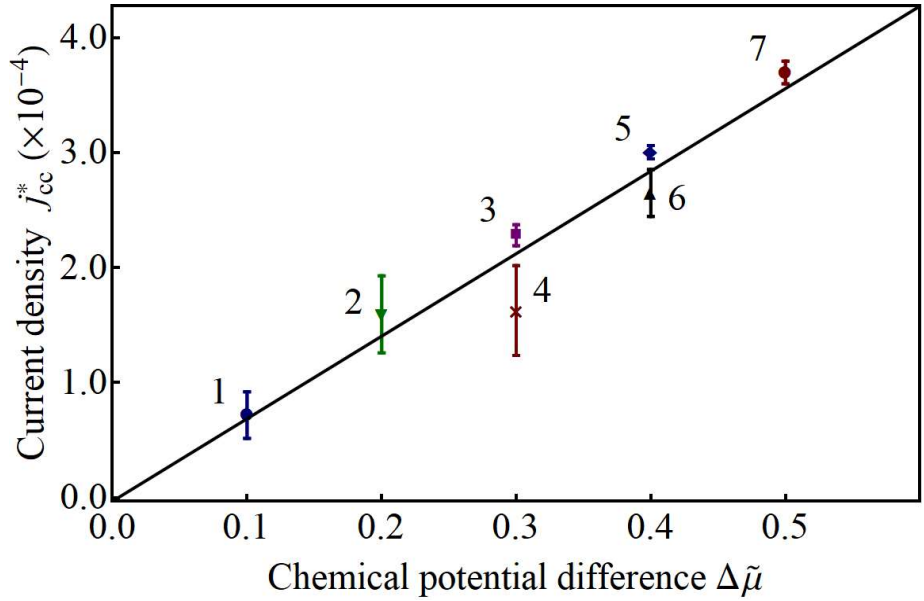

**Figure 11.** Dependence of the anode current density (Eq. 9) on the chemical potential difference $\Delta\tilde{\mu} = \Delta\tilde{\mu}_c - \Delta\tilde{\mu}_a$ during the discharging of the supercell shown in Fig. 5. The points correspond to the different values of the chemical potential difference in electrodes: 1 — $\Delta\tilde{\mu}_c = -0.35$, $\Delta\tilde{\mu}_a = -0.45$, 2 — $\Delta\tilde{\mu}_c = -0.3$, $\Delta\tilde{\mu}_a = -0.5$, 3 — $\Delta\tilde{\mu}_c = -0.2$, $\Delta\tilde{\mu}_a = -0.5$, 4 — $\Delta\tilde{\mu}_c = -0.2$, $\Delta\tilde{\mu}_a = -0.5$, 5 — $\Delta\tilde{\mu}_c = -0.1$, $\Delta\tilde{\mu}_a = -0.5$, 6 — $\Delta\tilde{\mu}_c = -0.2$, $\Delta\tilde{\mu}_a = -0.6$, 7 — $\Delta\tilde{\mu}_c = -0.2$, $\Delta\tilde{\mu}_a = -0.7$. The points are defined as shown in Fig.10. The point 4 shows the simulation results for the supercell with size of 625×128×128 (see Fig. S4 and S5), other points for the supercell with size of 350×128×128 (see Fig. 5).

## 6. Conclusion

The microstructure of the active electrode material determines to a large extent LIB performance[34,35]. In addition to the obvious aspects (active surface area, ion diffusion, and electrical conduction), it is necessary to take into account the dependence on morphology for processes such as phase separation, redistribution of electronic charge, effects of nonlinear diffusion, formation of grain boundaries, redistribution of elastic stresses, etc.

In this study, we have developed the 3D phase-field model of intercalation and transport of ions in lithium-ion batteries with realistic porous nanostructured electrodes. The model is based on the assumption that intercalation and transport occur at the constant value of overpotential that is related to the process of interfacial electrochemical reaction. We introduce the spatial dependence of dynamic and interaction parameters related to different properties of materials of anode, cathode and electrolyte. Also, we take into account the none-spherical form of electrode particles and their optional size distribution function. The spatial distribution of the electrode particles is generated by the special stochastic algorithm providing realistic porous electrodes with a given particle size distribution that can be derived experimentally. The model naturally describes the percolative identity of intercalant transport in interconnected electrode particles and electrolyte forming the network of the diffusion channels characterized by the different values of the species mobility.

In simulations, we have made some simplifications. Particularly, we have neglected the change in the volume of electrode particles during intercalation or extraction and have not taken into account contribution of the stresses and formation of defects. The goal of this study was to elaborate a phase-field model for a complete lithium-ion battery with realistic porous electrodes and to estimate the degree of deviation from the diffusion models commonly employed to predict battery properties. The obtained simulation results can contribute to better understanding of lithium transport processes in LIB essential for further improvement of the phase-field model.

An important feature of the model is that the electrochemical reaction is naturally determined by the chemical potential difference at the electrode-electrolyte interface eliminating the necessity to introduce the special boundary conditions at the interface. We have demonstrated that the change of the concentration inside the particle or in electrolyte can cause stopping of the intercalation process through the particle interface. Thus, we conclude that heterogeneity of intercalating atom distribution in electrolyte causes heterogeneity of insertion and extraction process in porous electrode. For a spherical electrode particle embedded into the uniform electrolyte phase, the insertion and extraction process can be described by the spherical concentration waves propagating from the particle interface to its center. This process agrees well with the classical approximation applied in the SP and P2D models of galvanostatic charging/discharging of a battery.

Basing on the developed phase-field model we have simulated the process of charging and discharging of a battery in 3D. The model is able to simulate intercalation and transport of solute in a battery. The electrode porosity can cause the inhomogeneous intercalation through the particle interface. We have demonstrated that in porous electrode insertion or extraction of intercalant in individual particle is heterogeneous over the particle interface and can occur through the mechanism of nucleation and growth or dissolution of the phase enriched with the intercalating atom. This effect shows the difference of the presented model from classical SP and P2D models implying equipartition of the current over the electrode particle interface. Also, insertion and extraction in porous electrodes can occur with the formation of the concentration front propagating from the electrode-separator boundary to the current collector. Formation of the concentration fronts have been observed in anode and cathode during the charging and discharging process. In the electrodes based on nanoporous and nanostructured materials, there are many alternative ways of diffusion, and the transport of ions is often percolative. Our model naturally takes into account the percolative identity of ion transfer with reactions at the electrode-electrolyte interface that is the fundamental difference from the SP and P2D models.

One of the basic assumptions of this model is the constancy of interfacial overpotential over all particles in electrodes. At the beginning of charging or discharging process the fast transient stage is observed. It can be related to short-range exchange of intercalant between electrode particles and electrolyte. Then transport between electrodes occurs with an almost constant value of current density and is explained by the long-range exchange of intercalant between electrodes. The current density remains constant until the possibility of insertion or extraction in electrode particles exhausts. After that we observe another short stage related to stopping of ion transport. Thus, the intercalation and transport of the solute at the constant value of overpotential can be related to galvanostatic discharging or charging of the battery. Also, we have found that an average value of the current density has a linear dependence on difference of overpotentials in anode and cathode.

The results highlight non-diffusive motion of the concentration front during charging or discharging, nonuniform intercalation flux over the surface of electrode particles and violation of the equipartition of the electric current density over the electrode surface. These effects are not taken into account by other models of lithium-ion batteries, that strongly limits their applicability.

**Acknowledgment**

This work is supported by the Russian Science Foundation (project No. 19-71-10063).

# Phase-field model of ion transport and intercalation in lithium-ion battery

Pavel E. L'vov[1,2*], Mikhail Yu. Tikhonchev[1], and Renat T. Sibatov[3]

[1] *Ulyanovsk State University, Ulyanovsk, Russia, 432017*
[2] *Institute of Nanotechnology of Microelectronics of the Russian Academy of Science, Moscow, Russia, 119991*
[3] *Moscow Institute of Physics and Technology, Dolgoprudny, Moscow, Russia, 141701*
[z] LvovPE@sv.uven.ru

**Algorithm of Electrode Formation**

A detailed description of the model used for simulation of the realistic electrode particle distribution can be found in Ref.S1. We have developed the special computer code to implement this model. Here, we briefly describe the main points of the modeling method. The construction of the electrode microstructure is carried out in several stages. The simulated electrode is a rectangular parallelepiped with the edges parallel to the Cartesian axes. The free boundary conditions are used along the X axis. The centers of the grains of the electrode material always lie inside the parallelepiped, but the grain itself can go beyond it. In contrast to the original approach, the boundary conditions along the Y and Z axes are periodic to satisfy the boundary conditions applied for solution of the CH equation (Eq.7). First, the system of nodes is randomly set in this parallelepiped. And then, the grains of the electrode material are built in these nodes. The spatial position of the nodes is determined by the Poisson process of the random distribution of points in space. Each node is randomly assigned with a linear grain size. At the initial stage, each grain is considered to be a ball with a center in the corresponding node. The ball radius is a random variable described by the gamma distribution with the parameters obtained from the experimental studies and given in paper Ref.S1. Then, on the set of pairs ($x_i$, $r_i$), where $x_i$ is the radius vector of the $i$-th node, $r_i$ is the sphere radius, the Laguerre mosaic[S2] is constructed. Next, the graph of the connectivity of the electrode grains is constructed. Each grain is described by a random Gaussian field on a sphere. The grain center is the center of mass of the corresponding cell of the Laguerre mosaic. The parameters of the Gaussian field are generated from the distribution functions given in Ref.S1 taking into account the connectivity of the grains. The code generates a file with coordinates of grain centers and the discrete sets of coordinates of points on the surface of each grain. The surface of grain can be approximated by a piecewise-linear closed surface, each element of which is a triangle. An example of a simulated grain structure of an electrode is shown in Fig. S1.

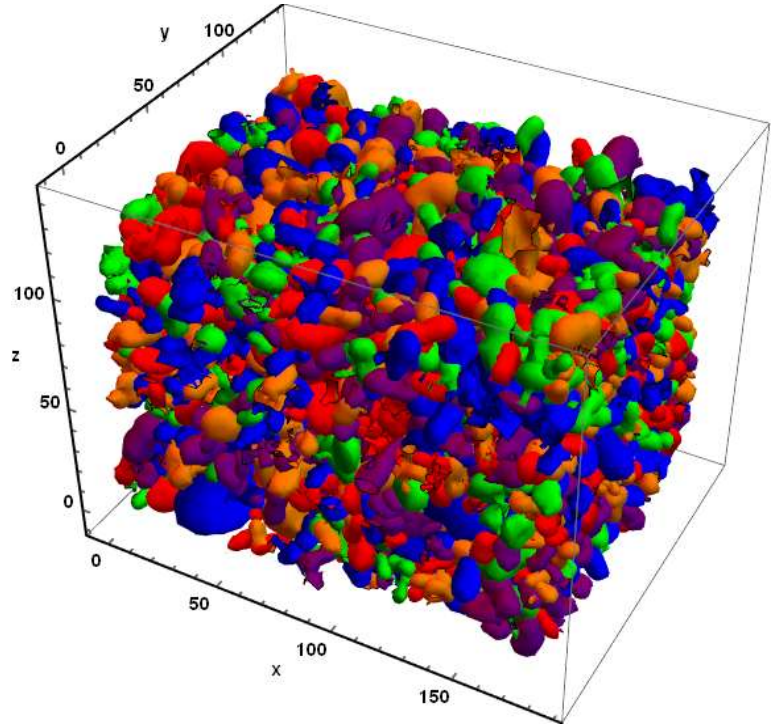

**Figure S1.** An example of the simulated microstructure of electrode with an average size of particles $\langle R \rangle = 5.2$. The electrode particles are shown in different colors for better perception.

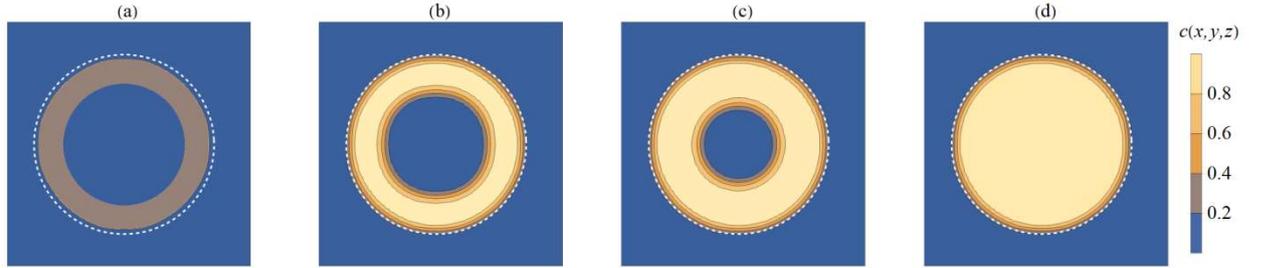

**Figure S2.** Dynamics of the concentration field during the solute insertion into individual electrode particle with radius $R = 20$ for $\Delta\tilde{\mu} = -0.6$. The distribution of the concentration field is given for $z^* = 64$, the size of the shown area is $48 \times 48$. Figures a-d describe different time points $t^*$: (a) - 100, (b) - 1100, (c) - 2100, (d) - 3200. Initial distribution of concentration field is uniform with equal concentration in the particle and electrolyte ($c_e = c_c = 0.07$). Dashed line shows the interface of the particle ($\xi = 0.5$).

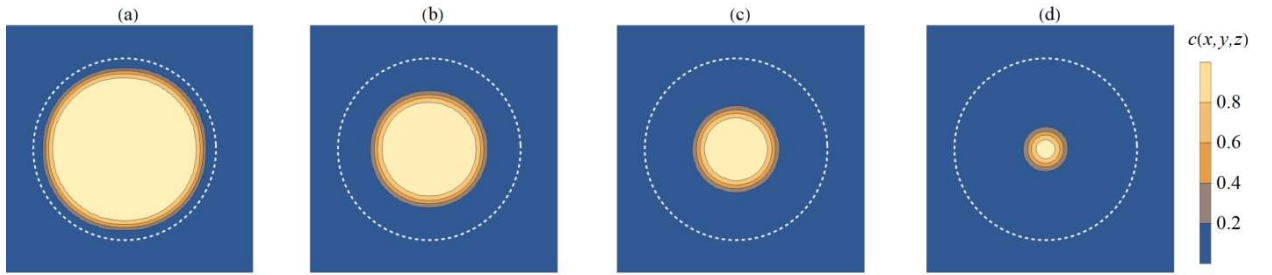

**Figure S3.** Dynamics of the concentration field during the extraction from individual electrode particle with radius $R^* = 20$ for $\Delta\tilde{\mu} = -0.2$. The distribution of the concentration field is given for $z^* = 64$, the size of the shown area is $48 \times 48$. Figures a-d describe different time points $t^*$: (a) - 500, (b) - 1600, (c) - 2600, (d) - 4300. Initial concentration in particles is $c_c = 0.9$ and in electrolyte $c_e = 0.07$. Dashed line shows the interface of the particle ($\xi = 0.5$).

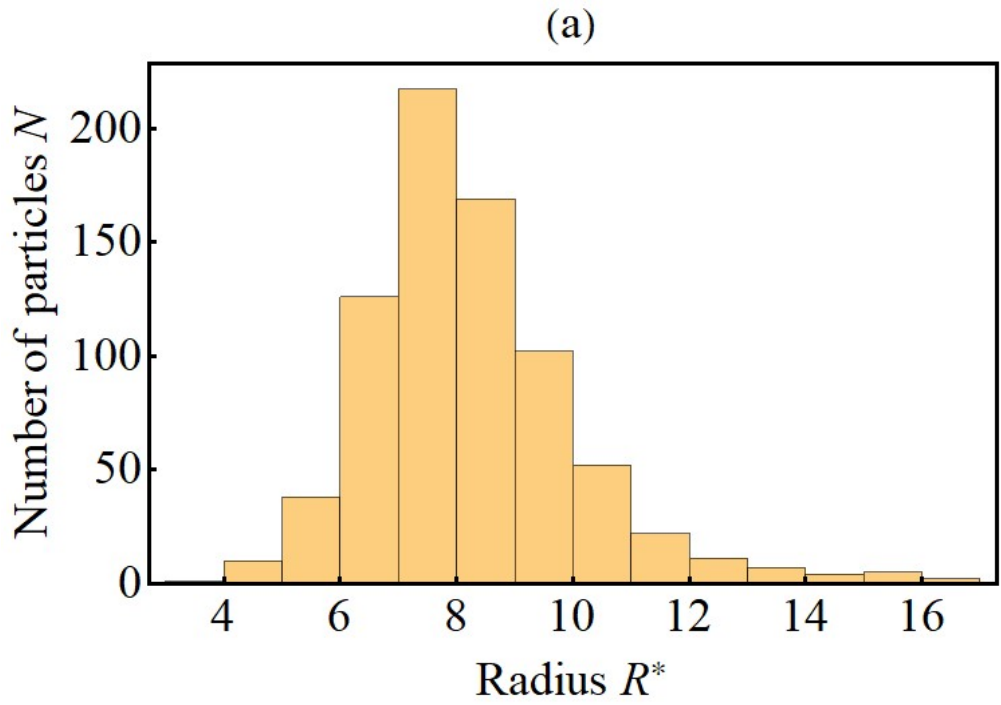

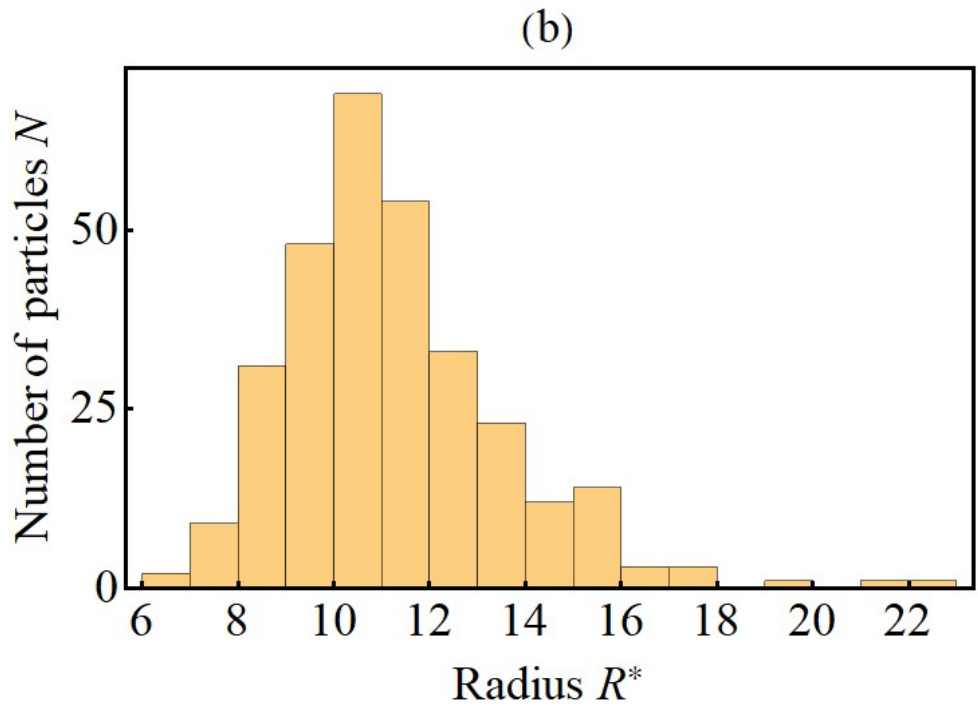

**Figure S4.** Size distribution function of electrode particles in cathode (figure a) and anode (figure b).

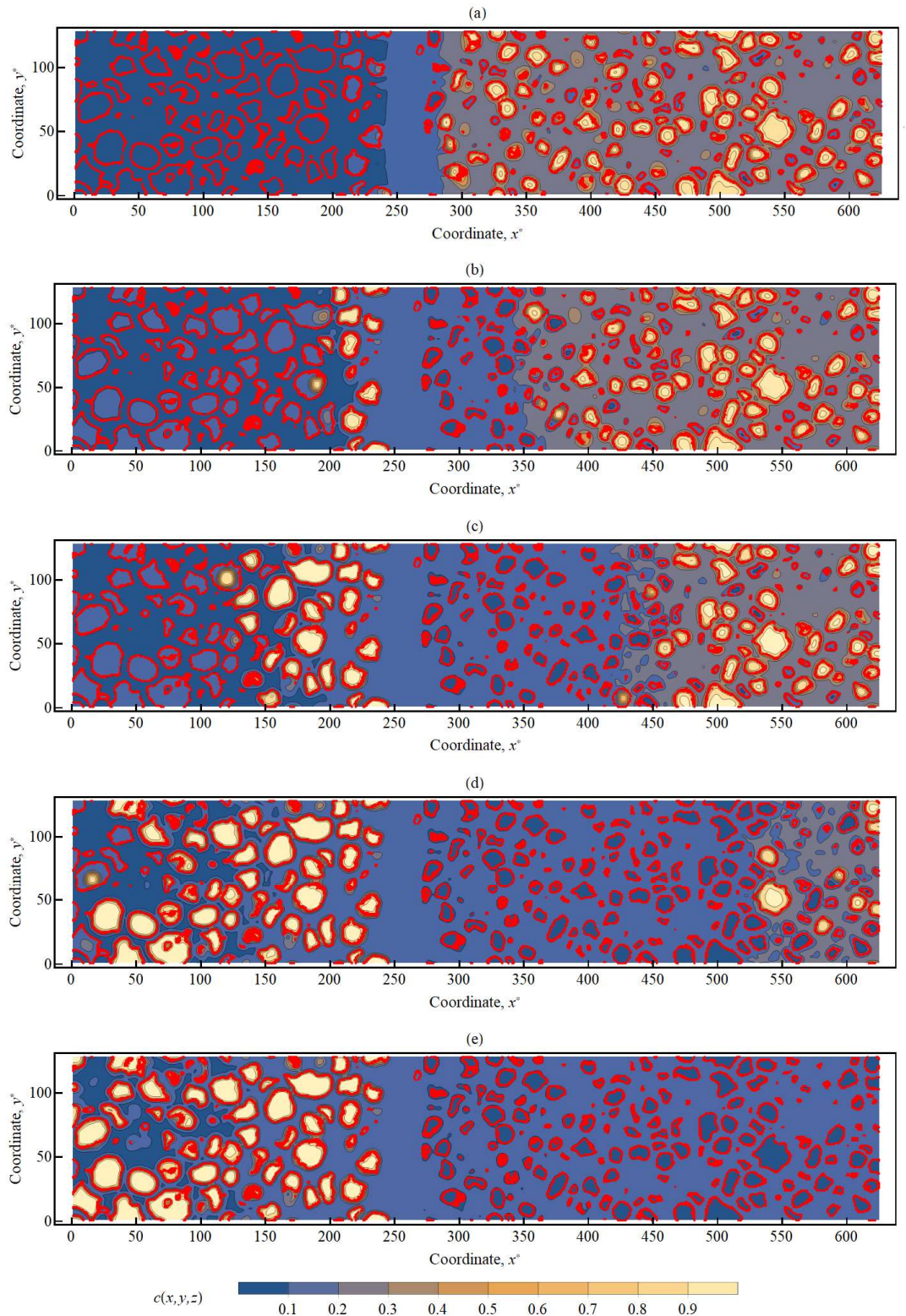

**Figure S5.** Transformation of the concentration field during the discharging of the battery with a size $620 \times 128 \times 128$. The figures correspond to different time points: (a) — $t = 500$, (b) — $t = 13500$, (c) — $t = 48500$, (d) — $t = 130500$, (e) — $t = 180000$. The simulation results are shown in the plane $z^* = 32$.

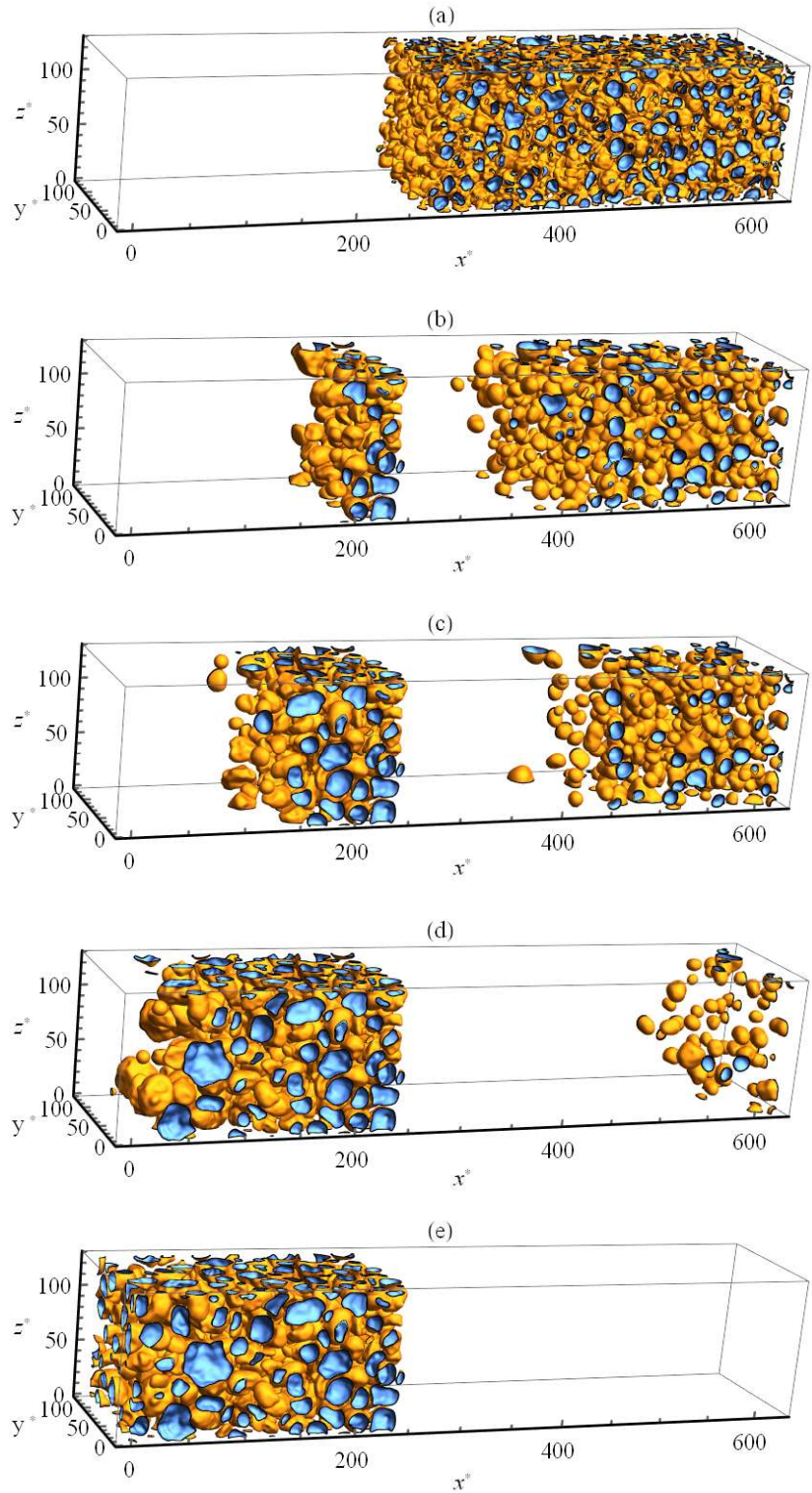

**Figure S6.** Dynamics of discharging of the battery having the size $V = 625 \times 128 \times 128$. The figures correspond to different time points: (a) — $t = 500$, (b) — $t = 13500$, (c) — $t = 48500$, (d) — $t = 130500$, (e) — $t = 180000$. The simulation results are shown by the surface of constant concentration $c = 0.5$.